\documentclass[%
 aip,
 jmp,%
 amsmath,amssymb,
reprint,%
]{revtex4-1}

\usepackage{graphicx}
\usepackage{dcolumn}
\usepackage{bm}

\begin{document}

\preprint{AIP/123-QED}

\title[On the Formation of Equilibrium Gels via a Macroscopic Bond Limitation]{On the Formation of Equilibrium Gels via a Macroscopic Bond Limitation}

\author{B. A. Lindquist}
\author{R. B. Jadrich}%
\author{D. J. Milliron}%
 \email{milliron@che.utexas.edu}
\author{T. M. Truskett} 
 \email{truskett@che.utexas.edu}
\affiliation{ 
McKetta Department of Chemical Engineering, University of Texas at Austin, Austin, Texas 78712, USA
}%

\date{\today}

\begin{abstract}
Restricting the number of attractive physical ``bonds'' that can form between particles in a fluid suppresses the usual demixing phase transition to very low particle concentrations, allowing for the formation of open, percolated, and homogeneous states, aptly called equilibrium or ``empty'' gels. Most demonstrations of this concept have directly limited the microscopic particle valence via anisotropic (patchy) attractions; however, an alternative macroscopic valence limitation would be desirable for greater experimental tunability and responsiveness. One possibility, explored in this paper, is to employ primary particles with attractions mediated via a secondary species of linking particles. In such a system, the linker-to-primary particle ratio serves as a macroscopic control parameter for the average microscopic valence. We show that the phase behavior of such a system predicted by Wertheim's first order perturbation theory is consistent with equilibrium gel formation: the primary particle concentrations corresponding to the two-phase demixing transition are significantly suppressed at both low and high linker-to-primary particle ratios. Extensive molecular dynamics simulations validate these theoretical predictions but also reveal the presence of loops of bonded particles, which are neglected in the theory. Such loops cause densification and inhibit percolation, and hence the range of viable empty gel state conditions is somewhat reduced relative to the Wertheim theory predictions. 

\end{abstract}

\pacs{Valid PACS appear here}
\maketitle

\section{\label{sec:intro}Introduction}

Gels comprising porous networks of colloidal particles linked via short-ranged physical bonds in a liquid matrix fill an important niche in materials engineering due to their viscoelastic and shear stress-driven yielding properties at low particle volume fractions.~\cite{gels_overview} On long time scales they function as fragile solids (i.e., small deformations induce yielding), but eventual bond rearrangement ultimately allows for flow and healing. Typically, gelation is induced by a quench below the spinodal of the demixing phase transition, whereby a kinetically arrested percolated particle network is established that preempts full phase separation.~\cite{gels_1,gels_2,gels_3} The resultant gel is non-equilibrium in nature and slowly ages as it phase separates. The ability to create more stable gels with greater experimental tunability for use in a variety of applications remains an outstanding materials design challenge.

Remarkably, recent developments have led to a new class of non-aging gels characterized by percolated networks at very low particle packing fractions that both 1) maintain equilibrium and 2) remain homogeneous even at the low temperatures (or high bond strengths) required to induce gelation.~\cite{fundamental_1,fundamental_2,fundamental_3,reentrant_1,reentrant_2,reentrant_3} At the heart of so-called \emph{equilibrium gel} fabrication is restriction on the number of bonds that can form between particles, thereby limiting the energetic driving force for particle condensation. Computational and theoretical studies of patchy particles, where the bonding possibilities are limited by the number and placement of patches on each particle, have demonstrated that the window for phase separation is strongly suppressed to lower densities by decreasing the number of patches; at two patches and below, the two-phase regime vanishes entirely.~\cite{valence_effect_1,valence_effect_2,valence_effect_3} As a result, a percolated, homogeneous, and sub-critical regime is accessible even at low particle concentrations in these systems, conditions that favor formation of so-called ``empty'' or equilibrium gels.

These theoretical predictions stimulated a number of recent attempts to experimentally synthesize equilibrium gels. One recently discovered system exploits the effectively patchy nature of Lamponite colloidal clay particles (imbued by their inhomogeneous surface charge and shape anisotropy) to fabricate gels of high stability at low particle density.~\cite{lamponite_clay} DNA tetramers and trimers with complementary sticky ends have also been shown to display this behavior.~\cite{dna_exp,dna_exp2,dna_sim_1,dna_sim_2} Finally, a relatively simple molecule, the dipeptide Fmoc-diphenylalanine, has been identified as a potential empty gel former, where attractions are largely mediated by $\pi$-$\pi$ stacking interactions of the side chains.~\cite{fmoc} 

In the aforementioned systems, the extent of bonding is determined by a microscopic particle valence limitation. While there has been progress in tuning specific interactions between particles, particularly with DNA-coated colloids where sections of single-stranded DNA provide specific binding sites,~\cite{exp_patchy_coll,coll_DNA1,coll_DNA2} such a requirement of microscopic control can generally present practical difficulties in terms of fabrication and tunability. One possible mechanism put forward for modulating the valency in the absence of a strict microscopic control is to add effective repulsions between the colloids.~\cite{eff_repulsions} In this work, we explore an alternative strategy of employing a more flexible macroscopically tunable surrogate via the use of a secondary species to link primary particles together. Importantly, inclusion of an explicit linking species is a generally relevant strategy to modulate inter-particle interactions: linking, or bridging, has been exploited in a diverse array of experimental systems, including (1) polymer-colloid mixtures~\cite{polymer_stabilize, binary_bridging_1,binary_bridging_2,binary_bridging_3,more_reentrant}, (2) oppositely charged nano-microsphere mixtures~\cite{gel_melt_exp1,gel_melt_exp2}, and (3) ion-linked, inorganic nanocrystal mixtures.~\cite{ChaM_NC,other_inorgNC} Inherent to linker-mediated interactions is the possibility of re-entrant phase behavior with respect to the linker-to-primary concentration ratio, $\Gamma$. From a homogeneous fluid, increasing $\Gamma$ induces gelation or phase separation; however, further increases in $\Gamma$ can recover a homogeneous phase as the surfaces of the primary particles becomes saturated with linker, inhibiting inter-particle bridging. A theoretical framework developed by Liu and co-workers~\cite{binary_bridging_2} for a binary mixture of hard-core spheres with complementary attractions supports a thermodynamic interpretation by showing that the range in density of the spinodal first grows and then shrinks as $\Gamma$ is increased.~\cite{ChaM_NC} We explore this effect in greater microscopic detail here for our model system and discuss its relevance to equilibrium gel fabrication.

We explore the possibility of forming equilibrium gels in systems where the interparticle attractions are mediated by linking species with a simplified model system that is an extension of the traditional patchy particle fluid. We study a binary mixture of particles, one species with six patches and the other with two patches, where the attractions between patches are complementary--only attractive between patches on unlike particles. Such a model will allow us to simultaneously make contact to prior studies on patchy particles but also to the linker-gel literature: the six-patch species, which has been shown to have similar phase behavior to a uniformly attractive (non-patchy) system,~\cite{valence_effect_1,fundamental_3} can be regarded as the primary particle and the two-patch species as the linker. While a strict microscopic valence is formally utilized within this model, particularly for the linker, a species with two interaction sites is a much less restrictive requirement in a linker system than in a pure colloidal system. Many candidate molecular or coordination bonding species can be readily identified that are ditopic, i.e., can form precisely two bonds in order to connect two particles in the fluid (e.g., divalent ions). Moreover, recent computational work studied bi-functional DNA designed to link DNA tetramers through complementary sticky ends~\cite{dna_linker}. Alternatively, a linker with no strict microscopic valency limitation might only allow for two bonds to primary particles for steric reasons if there is significant size asymmetry.

The remainder of the paper is organized as follows. In Section~\ref{sec:methods} we establish the theoretical and simulation methods used in this work. In Section~\ref{sec:results}, we demonstrate that both theoretical and simulation results predict that mixtures with both low and high $\Gamma$ ratios can yield equilibrium gels. We characterize points within the homogeneous regime with simulation, with an emphasis on those features that are unique to the macroscopic valence limitation. Finally, we summarize the main attributes of linker gels as revealed by our analysis in Section~\ref{sec:conclusions}.

\section{\label{sec:methods}Methods}

Here we outline the theoretical and simulation approaches for studying the phase behavior and properties of a system where linking particles provide a macroscopic constraint on attractive bond formation. The theory is derived for a binary mixture of particles (labeled type \(\mathcal{P}\) and \(\mathcal{L}\)), each with an arbitrary number of sites (\(n_{\mathcal{P}}\) and \(n_{\mathcal{L}}\)). Generality will be useful for making some universal analytical predictions in the Appendix, but the main text will focus on the case where \(n_{\mathcal{P}}=6\) (primary species) and \(n_{\mathcal{L}}=2\) (linking agent). A 6-patch primary particle was chosen because this valency does not display phase behavior consistent with equilibrium gelation in a single component system.~\cite{valence_effect_1,fundamental_3} For simplicity, the primary particle and linker are of equal diameter \(\sigma\) in the model, though the theoretical results are largely insensitive to particle sizes.

\subsection{Wertheim Theory}
\label{subsec:methods_wertheim}

Studies of patchy particle fluids have been greatly facilitated by Wertheim's first order perturbation Theory (WT), which is both relatively accurate and computationally simple.~\cite{wertheim_1,wertheim_2,wertheim_3} WT follows from an exact graphical expansion and partial re-summation of the partition function considering graphs where (1) no two particles share more than one bond, (2) each patch can participate in a maximum of one bond, (3) bonding occurs in a tree-like hierarchy (i.e., no closed loops), and (4) all bonds on a particle are probabilistically uncorrelated--approximations which perform best at low to moderate densities and a small number of patches. Some of these restrictions have been relaxed for specific systems~\cite{augmented_1,augmented_2,augmented_3}; however, WT in its standard formulation is typically reasonably accurate~\cite{reentrant_1,reentrant_2,reentrant_3} while retaining simplicity. Through these assumptions, a closed analytical expression for the Helmholtz free energy per particle, \(a\equiv a(\rho,x_{\mathcal{P}},T)\), can be derived which, for a binary mixture of \(\mathcal{P}\) (primary) and \(\mathcal{L}\) (linker) particles, depends on the total number density, \(\rho\), mole fraction of species \(\mathcal{P}\), \(x_{\mathcal{P}}\) [note: \(x_{\mathcal{L}}\equiv x_{\mathcal{L}}(x_{\mathcal{P}})=1-x_{\mathcal{P}}\)] and the temperature, \(T\). Ultimately \(a\) comprises two additive terms
\begin{equation} \label{eqn:wertheim_free_energy}
a \equiv a_{\text{HS}}+a_{\text{B}}
\end{equation}
where \(a_{\text{HS}}\equiv a_{\text{HS}}(\rho,x_{\mathcal{P}},T)\) is the Helmholtz free energy per particle for a binary hard sphere (HS) fluid and \(a_{\text{B}}\equiv a_{\text{B}}(\rho,x_{\mathcal{P}},T)\) is the bonding contribution that incorporates the patchy interactions. The expression for \(a_{\text{HS}}\) can be arrived at via integration of an approximate HS compressibility factor, \(Z_{\text{HS}}\equiv P_{\text{HS}}(\rho,x_{\mathcal{P}})/\rho k_{B}T\), where \(P_{\text{HS}}(\rho,x_{\mathcal{P}})\) is the hard sphere pressure and \(k_{B}\) is Boltzmann's constant:
\begin{equation} \label{eqn:wertheim_free_energy_HS}
\dfrac{a_{\text{HS}}}{k_{B}T} \equiv \sum_{i=\mathcal{P},\mathcal{L}}^{} x_{i}\text{ln}x_{i} + \text{ln}\rho-1 + \int_{0}^{\rho} (Z_{\text{HS}}-1) \text{d}\ln q
\end{equation}
where \(q\) is a dummy density integration variable and the virial Percus-Yevick mixture form for \(Z_{\text{HS}}\).~\cite{hs_eos_1,hs_eos_2} The bonding contribution \(a_{\text{B}}\) can be written using the convenient notation of Chapman et al.~\cite{wertheim_1,wertheim_2} as
\begin{equation} \label{eqn:wertheim_free_energy_B}
\begin{split}
a_{\text{B}} \equiv  \sum_{i=\mathcal{P},\mathcal{L}}^{} x_{i}\bigg[n_{i}\text{ln}X_{i} -\dfrac{n_{i}X_{i}}{2}+\dfrac{n_{i}}{2}\bigg]
\end{split}
\end{equation}
where \(n_{i}\) is the number of patches on a particle of species \(i\), and \(X_{i}\equiv X_{i}(\rho,x_{\mathcal{P}},T)\) is the fraction of those patches that are not bonded. The latter quantity depends on state point by solution of two mass-balance, chemical reaction-like equations which have the following simple form due to non-interaction between patches of the same species
\begin{equation} \label{eqn:mass_balance1}
X_{i}=[1+n_{j}x_{j}\rho X_{j} \Delta_{\mathcal{P},\mathcal{L}}]^{-1},\ i\neq j
\end{equation}
where \(\Delta_{\mathcal{P},\mathcal{L}}\) is an effective bonded pair partition function. For short attraction ranges \(\Delta_{\mathcal{P},\mathcal{L}}\) can be approximated as
\begin{equation} \label{eqn:mass_balance2}
\Delta_{\mathcal{P},\mathcal{L}}\approx 4\pi g_{\text{HS}}(\sigma)[\exp(\epsilon_{\mathcal{P},\mathcal{L}}/k_{b}T)-1]K_{\mathcal{P},\mathcal{L}}(\alpha_{\mathcal{P},\mathcal{L}},\sigma)
\end{equation}
where \(\alpha_{\mathcal{P},\mathcal{L}}\), \(K_{\mathcal{P},\mathcal{L}}\) and \(\epsilon_{\mathcal{P},\mathcal{L}}\) are the square-well range, bond volume and attraction strength, respectively, between patches on different species, and $g_{\text{HS}}(\sigma)$ is the contact value of the hard-sphere radial distribution function (Percus-Yevick result is used~\cite{hs_eos_2}).

\subsection{Phase Behavior}
\label{subsec:methods_phases}

The analytical nature of WT allows one to easily obtain information about the stability limit of the homogeneous fluid (i.e., the spinodal), reached when the \emph{compositional} stability criterion
\begin{equation} \label{eqn:composition_spinodal}
\bigg(\dfrac{\partial P}{\partial \rho}\bigg)_{T,x_{\mathcal{P}}} \bigg(\dfrac{\partial^{2} a}{\partial x_{\mathcal{P}}^2}\bigg)_{T,\rho}-\dfrac{1}{\rho^2} \bigg(\dfrac{\partial P}{\partial x_{\mathcal{P}}}\bigg)_{T,\rho}^{2} \geq 0
\end{equation}
is satisfied by the equality.~\cite{critical_phenomena} Here, the pressure \(P\equiv \rho^{2} (\partial a/\partial \rho )_{T,x_{\mathcal{P}}}\) is obtained via differentiation of Eq.~\ref{eqn:wertheim_free_energy}. Upon approach to the spinodal, the fluid becomes unstable via macroscopic fluctuations in the composition (\(x_{\mathcal{P}}\)), and phase separation into two (or more) homogeneous phases spontaneously occurs (except at the critical point). 
Note that this compositional instability always preempts the typical {\em mechanical} stability criterion relevant for one-phase systems:~\cite{critical_phenomena}
\begin{equation} \label{eqn:mechanical_spinodal}
\bigg(\dfrac{\partial P}{\partial \rho}\bigg)_{T,x_{\mathcal{P}}} \geq 0
\end{equation}
Nonetheless, as we explore below in Section~\ref{sec:results}, the relative locations of the compositional and mechanical spinodal boundaries provide insight into key differences between the behaviors of binary linker-gel systems and single-component patchy particle fluids.  

To complement the information provided by the spinodal boundary, one can also determine the equilibrium densities and compositions of the resulting coexisting phases. In the Appendix, we describe how we analyze two-phase coexistence in this work by directly casting the problem as a convex free energy minimization.~\cite{critical_phenomena,thermo_stat_mech} As compared to the standard approach of solving the non-linear equations that ensure equality of pressure and chemical potential between phases, the present strategy has better numerical stability and can take advantage of a diverse array of available optimization tools.

\subsection{Percolation}
\label{subsec:methods_percolation}

Subject to the assumptions regarding bonding outlined in Section~\ref{subsec:methods_wertheim}, one can locate the percolation threshold using WT.~\cite{wt_perc} Within this framework, a seed particle is required, which we choose arbitrarily to be of the species \(\mathcal{P}\) (this choice is irrelevant for a large cluster near percolation). Thereafter, alternating levels of \(\mathcal{L} \rightarrow \mathcal{P} \rightarrow \mathcal{L} \rightarrow \mathcal{P} ...\) occur, as only cross bonding interactions exist in our model. Thus, one can view one full ``effective'' bond as a sequence of \(\mathcal{P} \rightarrow \mathcal{L} \rightarrow \mathcal{P}\) particles implying the recursion relation between the number of \(\mathcal{P}\) particles at the \(y-1\) level, \(Q_{y-1}\), and the \(y+1\) level, \(Q_{y+1}\),
\begin{equation} \label{eqn:particle_bond_network}
Q_{y+1}^{(\mathcal{P})}=Q_{y-1}^{(\mathcal{P})}(n_{\mathcal{P}}-1)p_{\mathcal{P} \rightarrow \mathcal{L}}(n_{\mathcal{L}}-1)p_{\mathcal{L} \rightarrow \mathcal{P}}
\end{equation}
where \(p_{i \rightarrow j}\) is the probability of a particle of species \(j\) being bound to a site of species \(i\); the term \(n_{i}-1\) appears since one site must already be bonded in the network before the ``propagation event''. Since only unlike species can bind, \(p_{i \rightarrow j}\) is equivalent to the probability that a site of species \(i\) is bonded. This fact, combined with the uncorrelated bonding assumption within WT, implies \(p_{\mathcal{P} \rightarrow \mathcal{L}}=1-X_{\mathcal{P}}\) and \(p_{\mathcal{L} \rightarrow \mathcal{P}}=1-X_{\mathcal{L}}\).

Calculating the percolation threshold requires determining when the propagation of effective \(\mathcal{P} \rightarrow \mathcal{L} \rightarrow \mathcal{P}\) bonds is unbounded. As can be seen from Eqn. \ref{eqn:particle_bond_network}, this can only occur if
\begin{equation} \label{eqn:percolation_criterion}
(n_{\mathcal{P}}-1)p_{\mathcal{P} \rightarrow \mathcal{L}}(n_{\mathcal{L}}-1)p_{\mathcal{L} \rightarrow \mathcal{P}} \geq 1
\end{equation}
otherwise \(Q_{y+1}<Q_{y-1}\) and the number of particles at each level progressively decreases, leading to eventual termination. The percolation threshold is thus equivalent to equality in Eqn.~\ref{eqn:percolation_criterion} and is readily determined from \(X_\mathcal{P}\) and \(X_{\mathcal{L}}\).

\subsection{Simulation}
\label{subsec:methods_simulation}

In order to test the WT predictions, we perform molecular dynamics (MD) simulations of a binary mixture of patchy particles where the particles are composed of a central sphere with smaller spheres (representing patches) rigidly embedded such that their centers lie on the surface of the large sphere. The positions of the patches relative to the large sphere are held fixed and maximally spaced, i.e., the two patches on the linker are collinear with the large particle center, and the six patches of the primary particle are arranged in an octahedral fashion. Small perturbations to these positions were added to thwart crystallization, as only homogeneous phases are considered in the WT calculations. The large spheres all interact via a 24-12 WCA (hard-core-like) repulsion with a diameter of $\sigma$,
\begin{equation} \label{eqn:wca_potential}
\begin{split}
u_{\text{WCA}}(r)\equiv &H(2^{1/12}\sigma-r) \\
& \times \epsilon \Bigg(4 \bigg[\bigg(\dfrac{\sigma}{r}\bigg)^{24}-\bigg(\dfrac{\sigma}{r}\bigg)^{12}\bigg]+1\Bigg)
\end{split}
\end{equation}
where \(H(x)\) is the Heaviside step function, \(r\) is the radial distance between their centers, and \(\epsilon\) is the repulsive energy scale.

In addition to the central core repulsion, patch-patch interactions are captured via an isotropic, softened square-well (or shoulder) between patch centers
\begin{equation} \label{eqn:patch_potential}
u_{\text{P}|i,j}(r)\equiv -\epsilon_{i,j} \text{exp} \Bigg[-\Bigg(\dfrac{r}{\alpha_{i,j}}\Bigg)^{12}\Bigg]
\end{equation}
To encode a primary-linker patch attraction of reasonably short range we set \(\alpha_{\mathcal{P},\mathcal{L}}=0.3\sigma\), and \(\epsilon_{\mathcal{P},\mathcal{L}}=\epsilon\). In order to make contact with WT predictions, we must enforce the one-bond-per-patch assumption in the simulations.~\cite{wertheim_1,wertheim_2,wertheim_3} Often this is done by selecting a sufficiently short range of attraction between patches such that double bonding is forbidden via steric constraints.~\cite{fundamental_1,fundamental_2,fundamental_3,reentrant_1,reentrant_2,reentrant_3,valence_effect_1,valence_effect_2,valence_effect_3} However, this requires a very short attraction range ($\alpha_{\mathcal{P},\mathcal{L}} \approx 0.1\sigma$), and so the simulations must be performed at extremely low temperatures to be consistent with WT predictions. Instead, we take a different strategy that is enabled because the binary mixture only has cross attractions. Between \emph{like} patches, we place a repulsive shoulder characterized by $\alpha_{\mathcal{P},\mathcal{P}}=\alpha_{\mathcal{L},\mathcal{L}}=0.45\sigma$ and $\epsilon_{\mathcal{P},\mathcal{P}}=\epsilon_{\mathcal{L},\mathcal{L}}=-2.0\epsilon$ so that a bond formed between unlike particles will repel all other patches (note: the patches do not interact with the large spheres.) Therefore, there are excluded volume effects between like patches in the simulations that are not present in the WT calculations; however, we expect that such effects, like the patches themselves, are relatively small. In order to minimize such excluded volume effects, the repulsive hump is not so large that double bonding is strictly forbidden. Rather, double bonding is highly suppressed because the bond volume for a second bond is significantly reduced relative to bonding at an unoccupied site. At the lowest temperature that we study (i.e., the worst-case scenario for the model), only $\le 3\%$ of particles are over coordinated. 

Simulations are performed in the NVT ensemble with the LAMMPS Molecular Dynamics Simulator (LAMMPS).~\cite{LAMMPS} The number of particles is chosen such that there are no less than 1000 particles of each species. The integration time step, $dt$, is 0.0002$\sqrt{\sigma^{2}m/\epsilon}$. The temperature is maintained via a Nos\'{e}-Hoover thermostat with a time constant of $\tau=100dt$. The simulations were carried out between $T=0.07-0.15$ in increments of 0.01 (for notational convenience, $T$ is expressed in units of $\epsilon/k_{B}$ throughout the whole text). Equilibration was performed with a single simulation cooled via simulated annealing to $T=0.15$, where the temperature was then decreased by 0.01 over 5,000,000 MD steps; subsequently, another 5,000,000 equilibration steps at constant $T$ were carried out before cooling again. From these initial configurations, production runs of 10,000,000 MD steps were performed. It is very difficult to remain ergodic in the two-phase regime, and so we are mindful that such state points may well be kinetically trapped before full phase separation occurs (in as much is possible in a single finite sized box). However, within the homogeneous regime, the lack of evolution in the energy and properties indicates that the state points are well equilibrated. Yet, sampling at the lowest temperatures may be limited due to slow dynamics, a necessary trade-off with the large area of phase space explored via simulation in this work.

From the simulation data, clusters are identified in the usual way, with particles that are either (1) directly bonded or (2) connected via a pathway of bonded particles classified as a single cluster.~\cite{CSD_1,CSD_2} Bonds are identified between unlike particles that possess patches whose centers are within $\alpha_{\mathcal{P},\mathcal{L}}$ of one another. Percolation is identified when one (or more) cluster(s) in a snapshot spans the entire simulation box, as determined by an inability to perform a coordinate transformation for which the minimum image convention is not required to properly identify all bonded neighbors in a given cluster.

\section{Results and discussion}
\label{sec:results}
As described in the Introduction, one way to enable equilibrium gel formation is to reduce the concentrations corresponding to the two-phase demixing region so that a system at relatively low packing fraction ($\eta=\pi \sigma^{3} N/6V$) remains homogeneous even at low temperatures (i.e., strong bonding attractions) where kinetic arrest can occur. To determine whether the phase behavior of our model is consistent with equilibrium gel formation, we calculate the binodal and spinodal boundaries, as well as the percolation thresholds, at a variety of ratios of linker-to-primary particles ($\Gamma$). In the following subsections, we first use Wertheim Theory (WT) to explore phase space and then simulate some selected conditions to validate the WT predictions. 

\subsection{Theory}
\label{subsec:theory_results}
Percolation, coupled with kinetic arrest, is required to form a self-supporting gel. For a binary system with complementary attractions such as the one studied here, it is intuitive that some values of $\Gamma$ (i.e., if either species is in large excess) will not allow for a percolated network to form. The strategy outlined in Section~\ref{subsec:methods_percolation} and elaborated in the Appendix can be used to analytically predict which $\Gamma$ values will form a percolated network at $T=0$ for an arbitrary number of patches. For our model, where patch numbers $n_{\mathcal{P}} = 6$ and $n_{\mathcal{L}} = 2$ are used in Eq.~\ref{eqn:lower_bonding} and~\ref{eqn:upper_bonding}, we find that the lower limit of $\Gamma$  for which percolation can be achieved, $\Gamma_{\text{L}}$, is 0.6 and the corresponding upper limit, $\Gamma_{\text{U}}$, is 15.0. It is also possible to determine the ratio for which maximal connectivity occurs, $\Gamma_{\text{M}}$, which is 3 for our model (i.e., every patch participates in bonding at $T=0$ within the assumptions of WT). Below this threshold, some patches on primary particles will be nonbonded; similarly, above this ratio, there will be free patches on the linking species. In the Appendix, we also derive an expression for values of $\Gamma$ above and below $\Gamma_{\text{M}}$ which have equal propensities toward percolation at $T=0$, as defined by Eqn.~\ref{eqn:percolation_criterion}.

\begin{figure}[!htb]
  \includegraphics{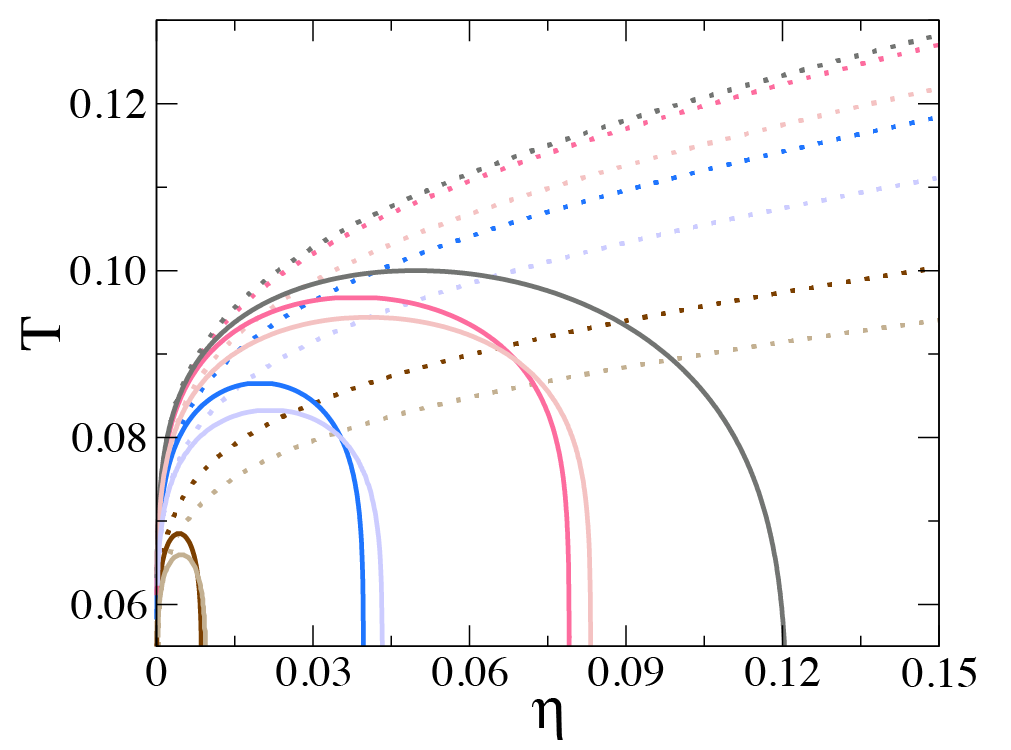}
  \caption{Spinodal boundaries (solid lines) and percolation thresholds (dotted lines) predicted by WT for equi-percolated pairs of $\Gamma$ (from left to right): 0.65 and 13.8, 0.9 and 10.0, 1.5 and 6.0, as well as $\Gamma=3.0$. The darker (lighter) curves correspond to conditions on the low- (high-)linker side.}
  \label{fgr:spinodal}
\end{figure}

These analytical predictions are confirmed by numerical WT calculations. Spinodal boundaries and percolation thresholds are not found by the WT calculations outside of the range given by $\Gamma_{\text{L}}$ and $\Gamma_{\text{U}}$. In Fig.~\ref{fgr:spinodal}, we show these boundaries as a function of $T$ and $\eta$ as calculated by WT for various values of $\Gamma$ within $[\Gamma_{\text{L}}$, $\Gamma_{\text{U}}]$. The pairs of $\Gamma$ values that are predicted by Eqn.~\ref{eqn:equi_percolation} to be equi-percolated (0.65 and 13.8; 0.9 and 10.0; 1.5 and 6.0) also have similar spinodal boundaries, though the percolation boundary associated with the high-linker side is depressed in $T$ relative to the low-linker side. These results imply re-entrant phase behavior (homogeneous $\rightarrow$ two-phase $\rightarrow$ homogeneous) with respect to increasing $\Gamma$ for an appropriate choice of $\eta$. In other words, an unstable region inside of a spinodal--conditions which would lead to spontaneous phase separation--only exists at intermediate values of $\Gamma$. This re-entrancy is consistent with the behavior of linker gels observed (1) theoretically by Liu and coworkers for a similar, but non-patchy, binary mixture with complementary attractions~\cite{binary_bridging_2,binary_bridging_3} and (2) experimentally by Singh, et al. for inorganic gels in which functionalized inorganic nanocrystals were linked by metal ions~\cite{ChaM_NC}. Furthermore, the suppression of the spinodal to low $\eta$ as $\Gamma_{\text{L}}$ and $\Gamma_{\text{U}}$ are approached, strongly suggests that it is possible to form equilibrium gels in these systems under such conditions because a low-density percolated fluid could remain homogeneous at very low temperatures (or high bond strengths).

\begin{figure}[!htb]
  \includegraphics{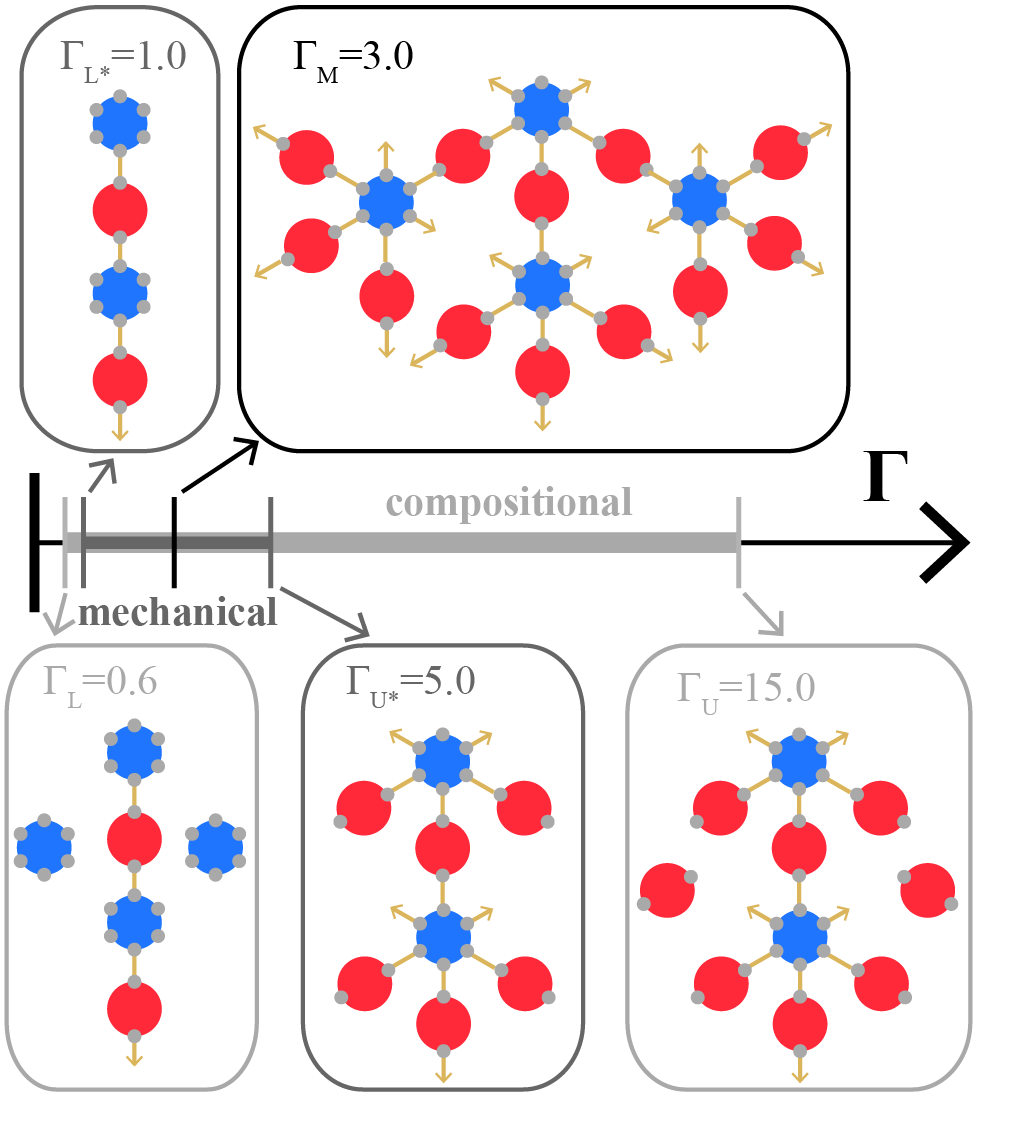}
  \caption{Summary of the ranges in $\Gamma$ where the spinodal boundary exists, using both the compositional and mechanical criteria, where the latter is denoted with an asterisk. Schematics (projected into two-dimensions) illustrating bonding patterns that are allowed at the critical values of $\Gamma$; only within the mechanical bounds can all particles participate in the network. Arrows outward indicate bond formation at that patch, even if the bonded particle is not depicted.}
  \label{fgr:scheme}
\end{figure}

The increase in the range of $\eta$ and $T$ for which spontaneous demixing will occur as $\Gamma$ increases from $\Gamma_{\text{L}}$ to $\Gamma_{\text{M}}$ bears similarity to what is observed in a single-component, patchy particle system as the number of patches is increased. At $\Gamma_{\text{M}}=3$, the spinodal region is maximally broad. Here, valence is not restricted by the macroscopic availability of linker since every patch can in principle participate in bonding. Above $\Gamma_{\text{M}}$, the primary particles become saturated with linker, limiting the number of bare patches on the primary particle. Therefore, the spinodal narrows; on a qualitative level, this narrowing from $\Gamma_{\text{M}}$ to $\Gamma_{\text{U}}$ is analogous to decreasing the patch number in a single-component, patchy particle system. While this analogy is appealing in its simplicity, there are some quantitative deviations that result from the explicit inclusion of the linker. For instance, in a single-component system, a minimum of two patches are required to form a percolated network, which corresponds to $\Gamma=1$ in our model (on average, each patch could donate one linker and accept one linker, yielding two bonds per primary particle). In fact, WT calculations using the mechanical criterion to determine the spinodal instability (see Section~\ref{subsec:methods_phases}) are in agreement with this prediction: the mechanical spinodal is not present for $\Gamma<1$. However, as Fig.~\ref{fgr:spinodal} demonstrates, when the compositional criterion is used, the spindal boundary does not vanish until $\Gamma<0.6$. Similarly on the high-linker side, the mechanical spinodal criterion predicts that the spinodal vanishes for $\Gamma>5.0$, where less than one bare patch per primary particle is available on average, but the compositional criterion predicts that the spinodal is present until $\Gamma=15.0$. Therefore, $\Gamma_{\text{L}}$ and $\Gamma_{\text{U}}$ imply microscopic bimodal character to the gel outside of the mechanical instability bounds: under these conditions, not all particles can be incorporated into a single percolated network. All of the preceding predictions are depicted schematically in Fig.~\ref{fgr:scheme}. 

\begin{figure}[!htb]
  \includegraphics{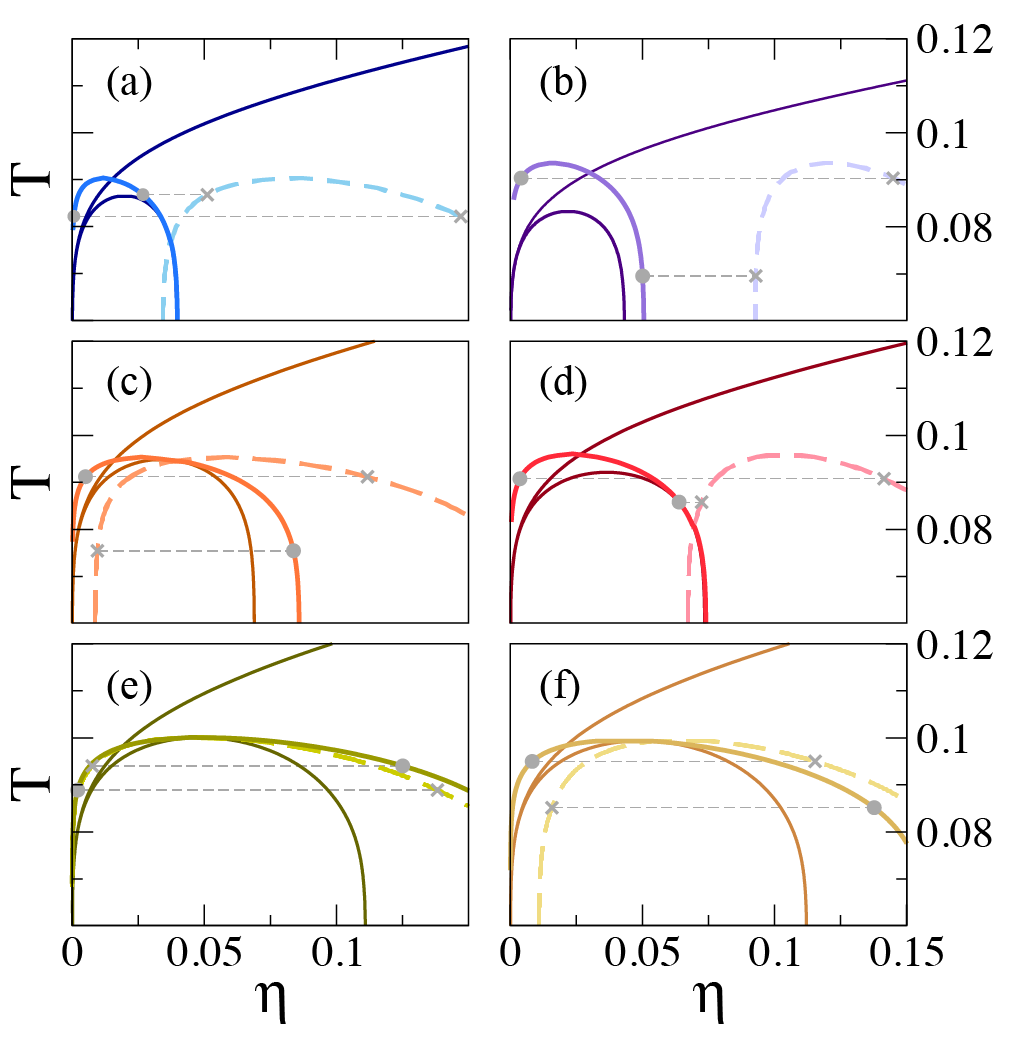}
  \caption{(a-f) Percolation threshold and spinodal boundary (thin dark lines), cloud curve (thick solid line), and shadow curve (thick dashed line) for $\Gamma=$0.9, 10.0, 1.3, 6.9, 2.5, and 3.6. At two densities for each $\Gamma$, gray dots on the cloud curve are connected by thin gray tie lines to the predicted density of the incipient phase, shown with an `x' on the shadow curve. Except for at the critical point, the shadow curve has a different composition than that of the bulk phase.}
  \label{fgr:phases}
\end{figure}

Clearly, the macroscopic nature of the linker constraint must be accounted for in a full description of the behavior of such linker-gel systems. Here, we also compute the phase coexistence (binodal) behavior, shown in Fig.~\ref{fgr:phases} together with the percolation threshold and the spinodal, for select values of $\Gamma$. The binodal, also called the {\em cloud curve}, represents the onset of phase separation--where an infinitesimal amount of a second incipient phase emerges in coexistence with the bulk phase.~\cite{cloud_shadow_1,cloud_shadow_2,binary_mixture_1,binary_mixture_2} The concentration of this incipient phase is given by the so-called {\em shadow curve}. Corresponding points on the cloud and shadow curves are joined by tie lines. Due to demixing, the composition of the incipient phase generally differs from $\Gamma$ of the bulk phase; that is, the shadow curve is a projection of state points onto the plane of fixed $\Gamma$. Therefore, the shadow curve \emph{does not} represent a literal phase boundary for a homogeneous system at a given $\Gamma$. The critical point, where the bulk and incipient phases are indistinguishable, corresponds to the point at which the cloud and shadow curves cross, which, unlike in a single component system, is not generally the maximum $T$ state point along the spinodal and binodal. However, the temperature at which physical bonds start to become long-lived and important can still be thought of as related to the highest temperature on the spinodal boundary, irrespective of where the critical point lies in the \(\rho, T\) plane.

The phase coexistence calculations shown in Fig.~\ref{fgr:phases} confirm that the binodal is suppressed to low densities as either $\Gamma_{\text{L}}$ or $\Gamma_{\text{U}}$ is approached; this phase behavior can be leveraged to fabricate equilibrium gels by sampling the adjacent, low \(T\) parameter space, analogously to what can be done in one-component patchy particle systems. Importantly, the theory also predicts that these anomalous, low-density equilibrium states are percolated within the relevant temperature range. Percolation is a necessary precursor to gelation, which will occur at some sub-percolation temperature whereby activated, slow bond-breaking dynamics will yield network rigidity over long, but finite, time scales. These long-lived, percolated networks are the equilibrium gel states. 

While pairs of equi-percolated points possess similar spinodal boundaries, they have markedly different phase coexistence behavior, with the shadow curve shifted to greater $\eta$ on the high-linker side. For instance, for $\Gamma=0.9$, the spinodal and cloud curve closely track and the critical point is at $T\approx 0.08$. For the corresponding value on the high-linker side, $\Gamma=10.0$, the spinodal and binodal are more separated and the shadow curve is so strongly shifted to high packing fractions that there is no critical point, i.e., the curves never touch in the plane of constant $\Gamma$. This can be understood because the fundamental unit of bonding on the high-linker side, a primary particle that is mostly saturated with linkers, is larger than that of the low-linker side, where it is a single primary particle and linker pair (see Fig.~\ref{fgr:scheme}). Therefore, the bonded network contains more particles on high-linker side under equi-percolated conditions, pushing the denser, bonded phase to higher overall volume fractions. As such, this effect is anticipated to be mitigated for smaller linkers, such as a metal-ion linked nanocrystal suspension~\cite{ChaM_NC}. 

The shadow curves lend insight into the relatively complex phase behavior that is possible within a multi-component system. In a single component system, the densities of the coexisting phases are given by points of equal temperature on the binodal. Therefore, in this context, equilibration within the two-phase regime yields a denser phase that is also an equilibrium gel if the binodal is suppressed to low densities. However, for a binary mixture, this strategy is complicated because phase separation results in compositional demixing, i.e., one must consult the cloud and shadow curves to know the densities of the resulting phases. As $\Gamma$ deviates from $\Gamma_{\text{M}}$, the tie lines in Fig.~\ref{fgr:phases}a,b,d indicate that the shadow curve is often shifted to higher densities relative to the cloud curve, particularly for low densities of the bulk phase. Additional complications due to compositional demixing occur as the system is quenched below the cloud curve, where a finite amount of the incipient phase exists, and therefore properties of the bulk phase are altered. These effects depend nontrivially on $T$, $\Gamma$, and $\eta$, but, on the whole, the coexistence calculations indicate that demixing tends to further densify the percolated phase by bringing $\Gamma$ closer to 3. 

Quasi-single-component behavior can be recovered near $\Gamma_{\text{M}}$, where the cloud and shadow curves have significant overlap (see Fig.~\ref{fgr:phases}c,f). However, in this regime, the binodal is too broad to generate a low-density equilibrium gel. For instance, at $\Gamma_{\text{M}}$, the packing fraction of the highly bonded phase is ~20\%. While this is significantly less dense than a single-component six-patch model due to the more severe microscopic valence restriction on the linker, this density would likely not be considered a low-density gel (though there is no rigorous cut-off). Therefore, most state points that phase separate are predicted to yield a relatively dense bonded phase at equilibrium, either via demixing or insufficient valence restriction, instead of a low-density gel.

In summary, these results indicate that while phase separation in linker-gel systems is rich and complex (compared to single-component, patchy particle systems) due to compositional demixing, equilibrium gel formation is indeed possible by preparing systems in the percolated homogeneous phase, which is readily accessible at low densities for certain values of $\Gamma$ where either primary \emph{or} linker particles are in excess. 

\subsection{Simulation}

\begin{figure}[!htb]
  \includegraphics[width=3.3in,keepaspectratio]{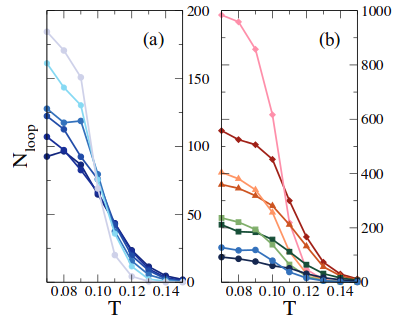}
  \caption{Number of loops per 1000 limiting particles as a function of temperature for (a) $\eta=$ 0.01 (lightest) to 0.06 (darkest) in increments of 0.01 at $\Gamma=$ 0.9 and (b) $\eta=$ 0.03 (lighter) and 0.10 (darker) at $\Gamma=$ 0.9 (circles), 1.1 (squares), 1.3 (triangles), and 6.9 (diamonds).}
  \label{fgr:loops}
\end{figure}

To validate the WT calculations, we performed simulations using a binary mixture of patchy primary and linker particles with attractions between unlike patches. We have aligned our simulation model with the assumptions of WT outlined in Section~\ref{subsec:methods_wertheim}: multi-bonding is highly suppressed via inclusion of repulsions between like patches, and patches are placed far apart on the particle surface so that 1) only one bond is possible between two particles and 2) to allow for relatively uncorrelated bonding. However, we cannot remove loops of bonded particles from simulations performed with pair potentials, nor is this necessarily desirable given that it is unclear whether such a model would have relevance to experimental systems.~\cite{loop_effects,valence_effect_3} As a result, we expect a major source of discrepancy between theory and simulation is due to loops present in the latter. As percolation is ultimately restricted by the limiting species, we report the number of loops, $N_{\text{loop}}$, in the simulations as loops per 1000 particles of the limiting species.~\footnote{The number of loops per cluster are calculated by $N_{\text{loop}}=N_{\text{bond}}-N_{\text{particle}}+1$~\cite{valence_effect_3}} In Fig.~\ref{fgr:loops}a, we show $N_{\text{loop}}$ as a function of $T$ between $\eta=$0.01$-$0.06 for $\Gamma=0.9$. For all $\eta$, loop formation is enhanced by lowering $T$ because the bonding probability is increased. When particles have a driving force to locally densify, as occurs upon network formation or phase separation, there is a reduced entropic penalty for looping. At the higher $T$, loop prevalence is greater for higher overall densities (a straightforward entropic effect); however, this trend is reversed for low $T$. This is a signature of phase separation within the finite box and can be understood via the phase coexistence calculations shown in Fig.~\ref{fgr:phases}a. For $\eta \leq 0.04$, where looping is most prevalent at low $T$, WT predicts phase separation. The shadow curve in Fig.~\ref{fgr:phases}a indicates that, as the bulk phase density decreases, the incipient phase densifies, favoring loop formation.

In Fig.~\ref{fgr:loops}b, we compare the number of loops at $\eta=$ 0.03 and 0.10 for $\Gamma=$ 0.9, 1.1, 1.3, and 6.9; $\eta=$ 0.03 is predicted to be in the two-phase regime for $T < 0.1$, and $\eta=$ 0.10 is predicted to be homogeneous for all $T$ at these compositions. For all $\Gamma$, the two values of $\eta$ exhibit the aforementioned crossover phenomenon in $N_{\text{loop}}$ as a function of $T$. For the low linker side, the number of loops increases as $\Gamma_{\text{M}}$ is approached and the valence is therefore less restricted; this is in accord with prior work on binary mixtures, where all patches were mutually attractive.~\cite{valence_effect_3} There are significantly more loops on the high linker side ($\Gamma=$ 6.9, equi-percolated with $\Gamma=$ 1.3); this can also be understood as a result of the fundamental bonding unit on the high-linker side being larger than on the low-linker side, thus making either the network (in the homogeneous regime) or the condensed phase (in the two-phase regime) more locally dense in the former. As with the shift in the range of the shadow curves to higher densities for high-linker conditions, this increased prevalence of loops may be dependent on the particle size asymmetry--an interesting question for future research.  

\begin{figure}[!htb]
  \includegraphics{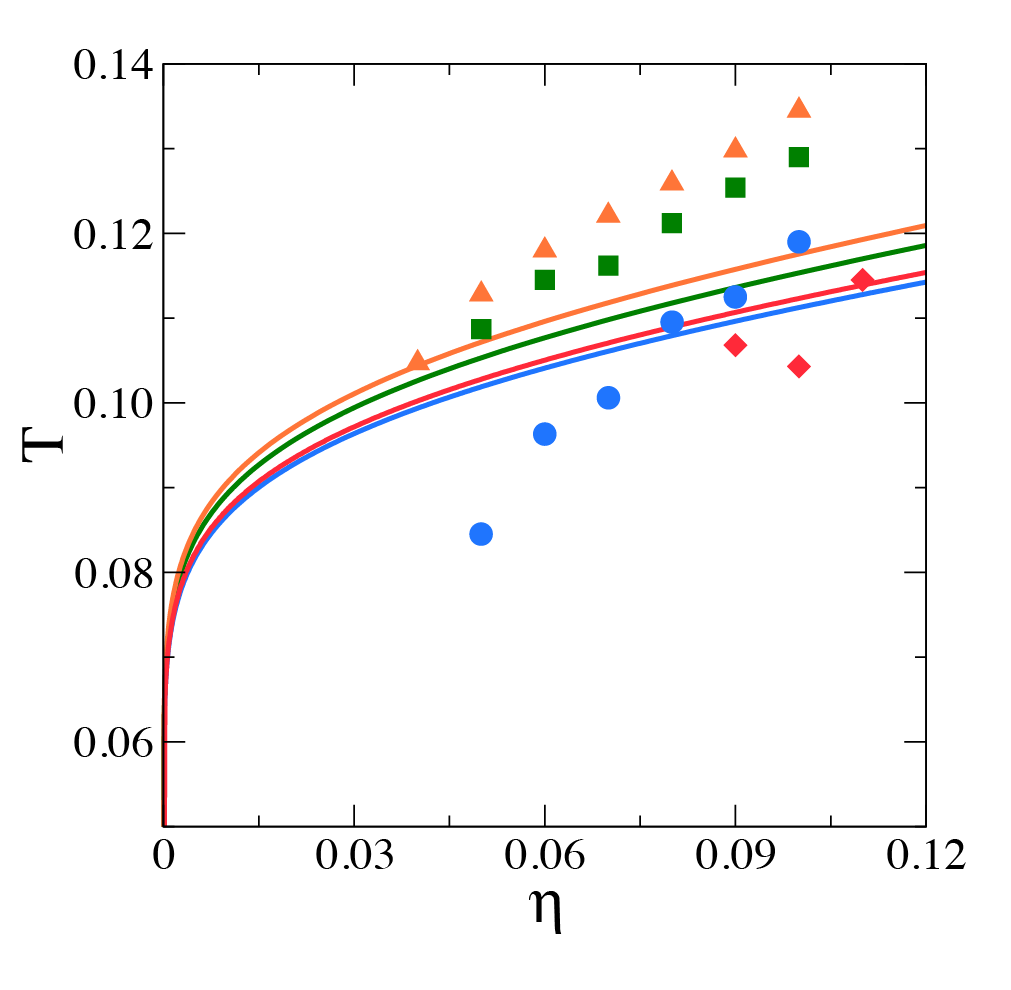}
  \caption{Percolation thresholds calculated by WT (lines) and simulation (symbols) for $\Gamma=$ 0.9 (circles), 1.1 (squares), 1.3 (triangles), and 6.9 (diamonds). The latter are determined by fitting the percentage of configurations containing at least one percolated cluster as a function of $T$ to a logit function and extracting the temperature where 50\% of snapshots are percolated from the fit.}
  \label{fgr:perc}
\end{figure}

Loops reduce the spatial extent of the network; therefore, we anticipate that they might suppress the percolation threshold. In Fig.~\ref{fgr:perc}, we compare the percolation thresholds predicted by theory and simulation for $\Gamma=$ 0.9, 1.1, 1.3, and 6.9. Only state points in the homogeneous regime are plotted. The mapping between the simulated temperature and temperature in WT is approximate, which likely accounts for the higher percolation temperatures observed in simulation for most state points. On the low linker side, the qualitative ordering of the thresholds with respect to $\Gamma$ is preserved, though $\Gamma=0.9$ appears more suppressed than might be anticipated based on $\Gamma=1.1, 1.3$, particularly at lower densities. In the simulation, the threshold for $\Gamma=6.9$ is lower in temperature than for $\Gamma=0.9$; WT predicts that these compositions have near identical boundaries (with the opposite ordering with respect to $T$).

Qualitative disagreement between theory and simulation is seen at $\Gamma$ values that are even further from $\Gamma_{\text{M}}$: several $\Gamma$ values outside of the range shown in Fig.~\ref{fgr:perc} do not percolate in the simulation. For $\Gamma=0.65$, a percolation threshold was not found for $T\geq0.07$, even up to $\eta=$ 0.20. The high linker side is even more affected, with no percolation threshold found for $\Gamma \geq 7.5$ (equi-percolated with $\Gamma=1.2$). This might be due to the increased loop formation on high-linker side, but also the high-linker side percolation thresholds are lower in $T$ than their counterparts on the low-linker side in the WT calculations. Even though there are fewer loops in absolute terms as compositions deviate from $\Gamma_{\text{M}}$, even a single loop precludes percolation at $\Gamma_{\text{L}}$ or $\Gamma_{\text{U}}$. As a result, the percolation thresholds at compositions differing the most from $\Gamma_{\text{M}}$ seem to be more sensitive to loop formation. It is also possible that the seemingly increased suppression of the simulated percolation thresholds as $\eta$ decreases, despite the presence of fewer loops at lower $\eta$ within the homogeneous regime, is due to similar reasons. 

\begin{figure}[!htb]
  \includegraphics{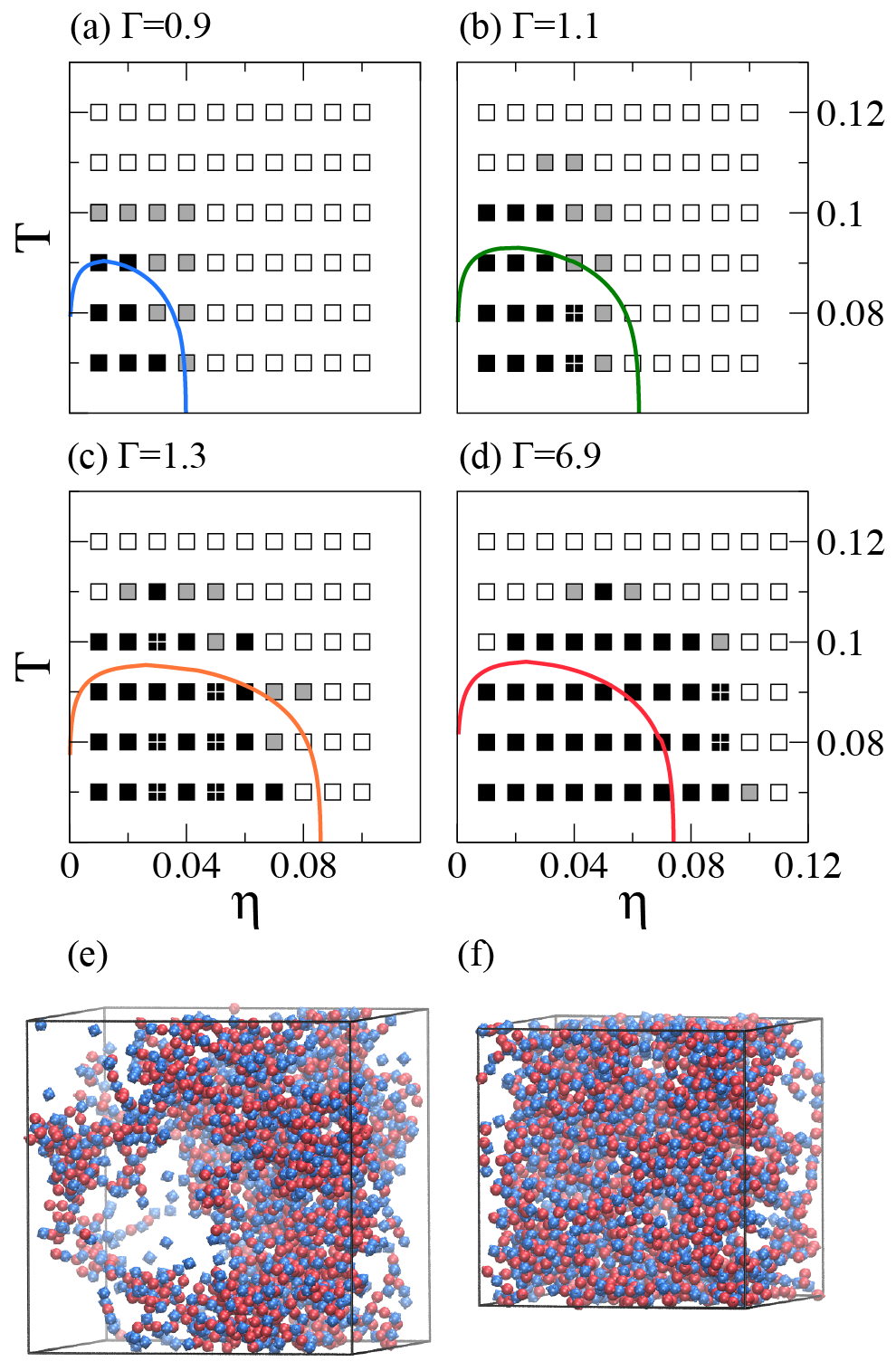}
  \caption{(a-d) Comparison of the binodal from WT and the phase behavior observed in simulation, where black squares denote phase separation as indicated by $S(k=0)>10$, black squares with white crosses indicate phase separation determined by a normalized standard deviation in $\eta$ being above 25, even though $S(k=0)<10$ (see the main text), gray squares indicate proximity to phase separation where $5<S(k=0)<10$, and white squares are homogeneous state points where $S(k=0)<5$ for $\Gamma=$ 0.9, 1.1, 1.3, and 6.9. Representative configuration of (e) a phase-separated state point ($\Gamma=1.3$, $\eta=0.05$, $T=0.07$) and (f) a homogeneous state point ($\Gamma=1.3$, $\eta=0.09$, $T=0.07$). Blue spheres are the primary species and red spheres are the linkers. Visualizations were created with VMD.~\cite{VMD}}
  \label{fgr:PSsim}
\end{figure}

While loops appear to reduce the range in $\Gamma$ for which percolation occurs, many $\Gamma$ values are percolated and therefore could potentially form an equilibrium gel if the binodal is sufficiently narrow. Here we compare the phase behavior observed via simulation to the binodals computed with WT. To detect phase separation in simulation, we monitor low-$k$ behavior of the structure factor, $S(k) \equiv \Big < N^{-1} \sum_{i,j}^{} \text{exp} [-i \mathbf{q} \cdot (\mathbf{r}_{i} - \mathbf{r}_{j} )] \Big > $, as a function of $T$. A divergence at $k=0$ indicates macrophase separation.~\cite{liquid_state_theory,fundamental_3} We perform this analysis for a grid of state points as shown in Fig~\ref{fgr:PSsim}a-d. We define phase separation as $S(k=0)\ge 10$, where the $k=0$ value is computed via linear extrapolation, and show such points as black squares in Fig~\ref{fgr:PSsim}. Points where $5<S(k=0)<10$ are denoted with gray squares, which can be interpreted as conditions that are near to a phase boundary. These criteria can be impacted by finite size effects: as $S(k=0)$ diverges, the correlation length grows, narrowing the peak about $k=0$.~\cite{liquid_state_theory} Since $S(k)$ can only be computed at wavelengths that fit in the box, the full behavior of the $k=0$ peak may not be captured if the correlation length is particularly large. Thus, as a cross-validation step, we consider the standard deviation in $\eta$ computed from eight distinct cubes, 10$\sigma$ in length, in a given configuration, averaged over the trajectory (similar to previous work~\cite{local_fluct}). We then normalize this value by the standard deviation in an analogous hard sphere simulation (in practice, we turn the patch interactions off). This metric rises sharply as phase separation occurs, but can be affected by the percolation threshold, as well as $\Gamma$ and $\eta$. However, as a general rule, we find that phase separation occurs when the standard deviation in $\eta$ is $15-25$ times greater than in the hard sphere analogue. Therefore, if the normalized standard deviation is greater than 25, then we classify the state points as phase separated, even if $S(k=0)\le 10$. Such points are denoted by a black square with a white cross. When these points appear near the phase boundary, they indicate that these state points are borderline; when they appear within a phase separated window it most likely means that the correlation length is so large that the computed $S(k)$ is not capturing the divergence at $k=0$ for the box size. See, for example, Fig.~\ref{fgr:PSsim}e.

On the whole, Fig.~\ref{fgr:PSsim} shows that the general trends predicted by WT are recovered. The high-linker side is somewhat broader than predicted, whereas the low-linker side deviates towards a narrower two-phase window, but in both cases the two-phase window terminates at reasonably low densities. Also, as with the percolation thresholds, the simulated two-phase regimes consistently extend higher in temperature than the WT predictions. 

\begin{figure}[!htb]
  \includegraphics{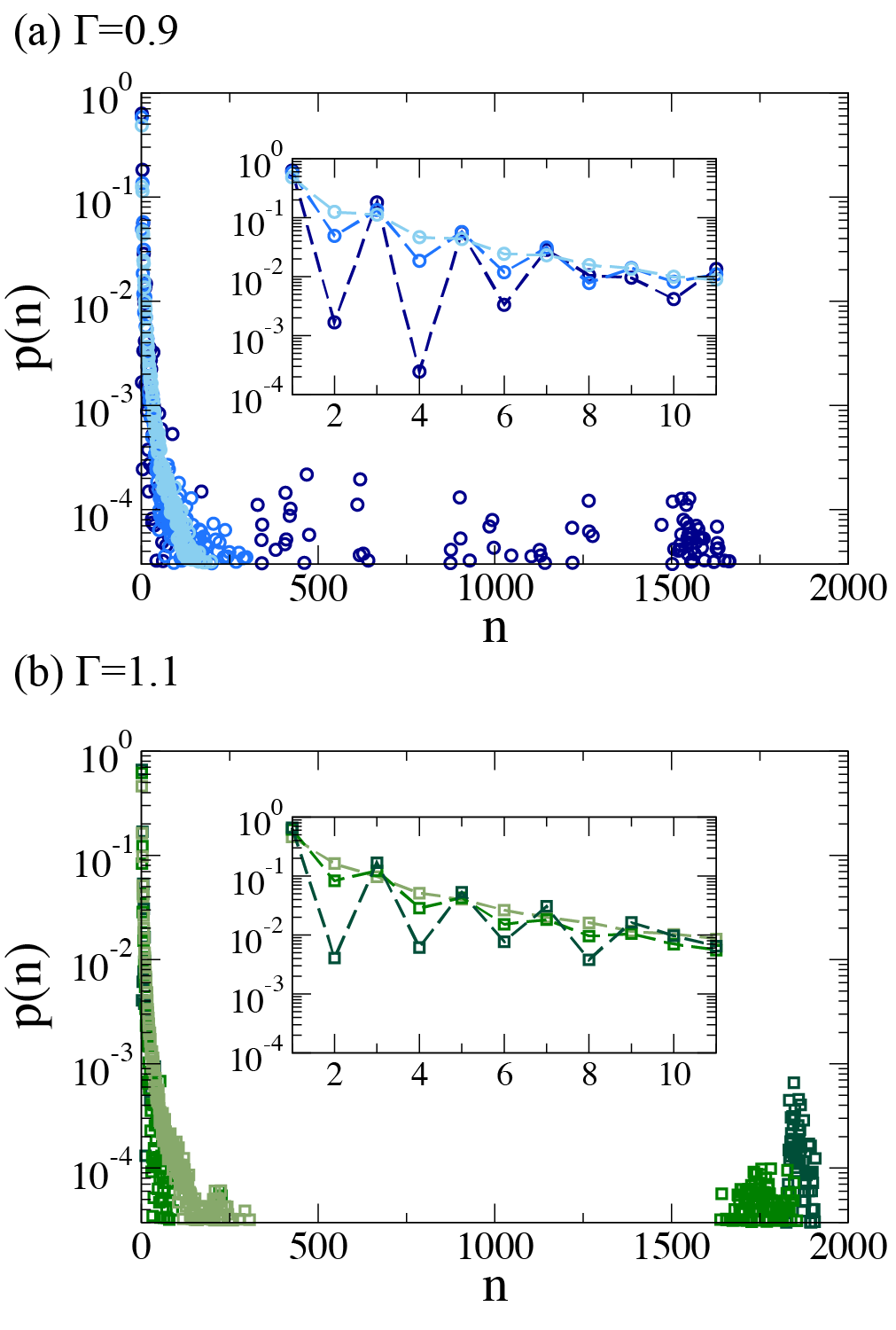}
  \caption{Cluster size distributions (CSDs) for $T=$ 0.07, 0.10, and 0.12, from darkest to lightest for (a) $\Gamma=$ 0.9 and (b) $\Gamma=$ 1.1. (Insets) Enhanced view of CSDs for small clusters.}
  \label{fgr:csds}
\end{figure}

We have confirmed with simulation that a relatively low-density homogeneous regime exists for many values of $\Gamma$; see for instance, Fig.~\ref{fgr:PSsim}f. The structures of the low temperature simulations are in keeping with the fundamental picture described by Fig.~\ref{fgr:scheme}: a percolated network in coexistence with monomer of the excess species, as well as small ``strings'' and loops. The cluster size distribution, or CSD~\cite{CSD_1,CSD_2}, is shown in Fig.~\ref{fgr:csds} for $\Gamma=$ 0.9 and 1.1 at $\eta=0.07$; both conditions are homogeneous at all values of $T$. As discussed in Section~\ref{subsec:theory_results}, for $\Gamma<1.0$, it is not possible to incorporate all particles into a single network. As a result, while the system is percolated at $\Gamma=0.9$ and $T=0.07$, the space spanning network contains only $\approx$ 2/3 of the total particles and is not the only cluster of significant size in the simulation. By contrast, when $\Gamma=1.1$, the percolated cluster contains $> 90\%$ of the particles in the simulation at $T=0.07$. The higher temperatures ($T=0.10, 0.12$) are somewhat less sensitive to composition, though $\Gamma=1.1$ clearly has a higher percolation threshold temperature at this density (consistent with Fig.~\ref{fgr:perc}). The insets show the behavior of the CSD for smaller cluster sizes. As the system is cooled, a ringing emerges, with odd-numbered cluster sizes dominating for small cluster size ($n$). As the temperature is lowered, the limiting species becomes nearly completely bonded, so effectively every cluster is capped by the excess, in this case the primary, species. This ringing dies off around $n=8-11$ because of looping: a loop removes one excess particle from the cluster, allowing for the limiting species to be fully bonded in a cluster with even $n$.  

\begin{figure}[!htb]
  \includegraphics{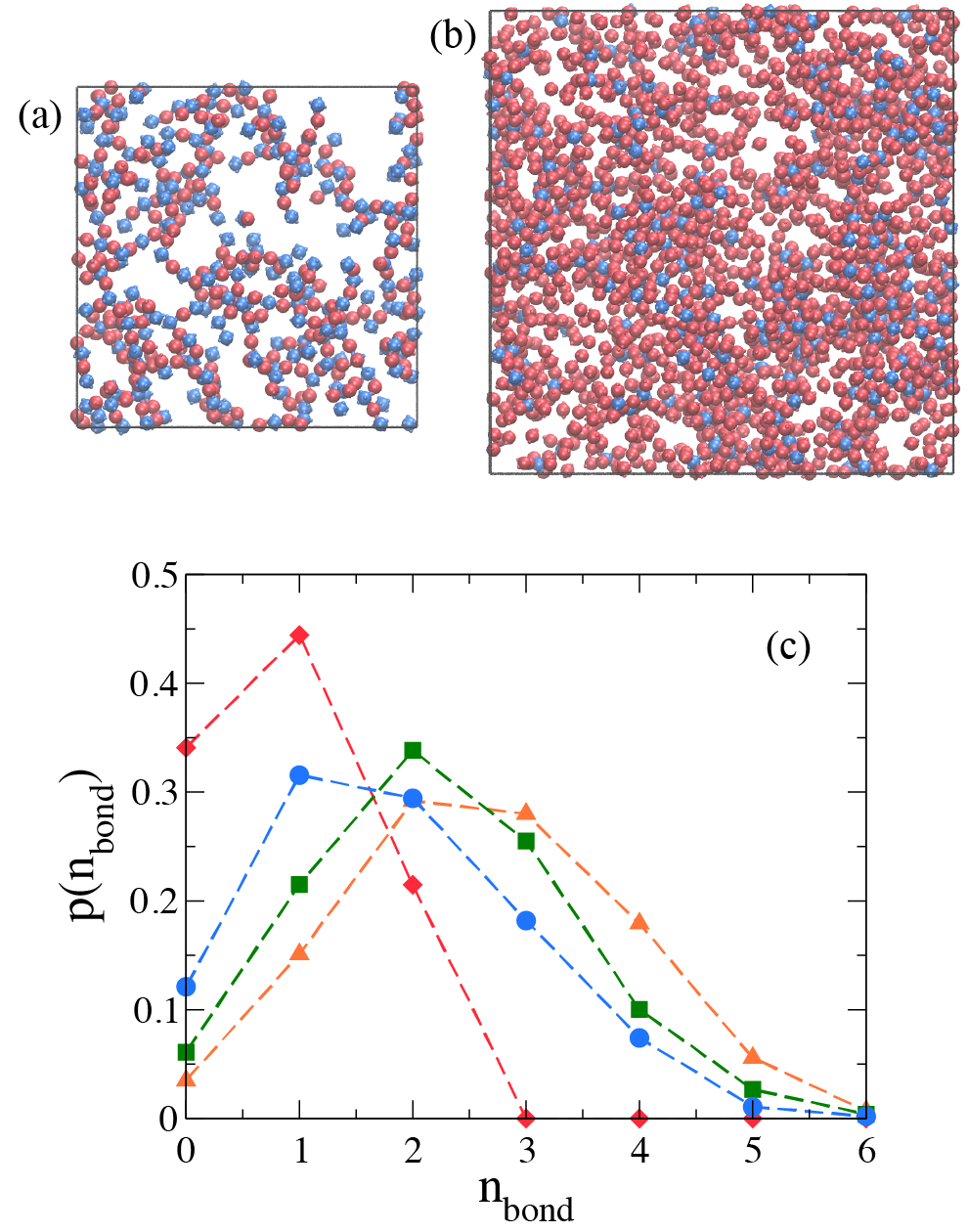}
  \caption{At (a) $\Gamma=1.3$ and $\eta=0.09$ and (b) $\Gamma=6.9$ and $\eta=0.11$, a 5$\sigma$ thick slab from a representative configuration at $T=0.07$. Blue spheres are the primary species and red spheres are the linkers. Visualizations were created with VMD.~\cite{VMD} (c) Also at $T=0.07$, probability of number of bonds per excess species for $\Gamma=$ 0.9 (circles), 1.1 (squares), and 1.3 (triangles) at $\eta=0.08$, and $\Gamma=$ 6.9 (diamonds) at $\eta=0.11$.}
  \label{fgr:nNN}
\end{figure}

Visualization of the lowest temperature conditions studied here reveals a reasonably stringy structure of alternating particle species. These string-like components are similar in spirit to other patchy systems that form low-density homogeneous states~\cite{reentrant_1,reentrant_2,reentrant_3,valence_effect_1,valence_effect_2,valence_effect_3}, but the number of bonds per particle is, of course, not as strictly limited. Fig.~\ref{fgr:nNN}a,b show 5$\sigma$ thick slabs from configurations from the low ($\Gamma=1.3$) and high ($\Gamma=6.9$) equi-percolated linker compositions, respectively. On the high-linker side, there are many free linkers in coexistence with the percolated network; the network can be identified in the image from the correlations in the primary (blue) particles. The low-linker configuration appears more open in nature owing in part to reduced numbers of totally unbound particles of the excess species ($\eta$ is also somewhat smaller in Fig.~\ref{fgr:nNN}a than in Fig.~\ref{fgr:nNN}b). The distribution of valencies is quantified in Fig.~\ref{fgr:nNN}c, where the probability of a given valence for the excess species is plotted at $T=0.07$. (At this temperature, the vast majority ($> 96\%$) of the limiting species is fully bonded.) On the low-linker side, it is not uncommon for a primary particle to only bond to two linkers, but $n_{\text{bond}}=1$ and 3 are also relatively probable. When $n_{\text{bond}}=1$, this indicates a dead-end in the network structure, and $n_{\text{bond}}=3$ is a branch point. Prior work has shown for a binary system with 2 and 3 patches (where all patches attract) that increasing the latter densifies the system significantly.~\cite{valence_effect_1,valence_effect_2,valence_effect_3} The presence of dead-ends and branch points, some of the latter with $n_{\text{bond}}=4$ and 5, will furnish network that is more spatial heterogeneous than a network composed of microscopically restricted particles. On the high-linker side, there are many dead-ends, where the linkers have saturated the surface of the limiting primary particles, but there are also many linkers free in solution ($n_{\text{bond}}=0$), as required for $\Gamma>5$. Only $\approx 20\%$ of the linkers actually contribute to the percolated network in this case. 

\section{Conclusions}
\label{sec:conclusions}

We have demonstrated the extent to which a macroscopically tunable bond valence control parameter--the linker-to-primary particle ratio ($\Gamma$)--can be leveraged in a model binary colloidal system with complementary attractions to enable formation of equilibrium gels. This represents an important, practical extension to existing work on single-component patchy particle fluids in which valence is controlled by tuning the number of patchy particle sites. Experimentally achieving microscopic tunability in these latter systems is a significant challenge, even for the most advanced nanoparticle and polymer colloidal synthesis techniques--making the possibility of an alternative macroscopic control appealing. 

Using Wertheim theory, our linker-gel model system is predicted to exhibit re-entrant phase and percolation behavior with respect to $\Gamma$. Only in a finite range of \(\Gamma\), namely, \(\Gamma_{\text{L}} < \Gamma < \Gamma_{\text{U}}\) are a two-phase region and percolation threshold present. In between the upper and lower bounds is a state of saturated bonding at \(\Gamma_{\text{M}}\). At this point the two-phase window is maximally broad in density and deviating towards either \(\Gamma_{\text{L}}\) or \(\Gamma_{\text{U}}\) is accompanied by a continuous suppression of the two-phase region to low density. This effect can be leveraged to access a regime of low density, percolated, homogeneous state points which at sufficiently low temperatures could form equilibrium gels.  

Using molecular dynamics simulation, we exhaustively tested the Wertheim theory predictions; overall, good qualitative agreement was found. However, network loop formation--which is neglected in Wertheim theory--generates important quantitative deviations. Loops shrink the range (\(\Gamma_{\text{L}}\leftrightarrow\Gamma_{\text{U}}\)) over which percolation is observed, the practical consequence of which is to limit the predicted state conditions available for forming empty gels. In this work, we have explored the case where the primary species and linker are equal in diameter, but the critical values in $\Gamma$ are independent of size ratio. Therefore, the findings of this work related to re-entrant phase behavior and percolation should be applicable to size asymmetric mixtures, though the packing fractions will be rescaled. 

While the linker model employed here formally possesses a strict microscopic valence restriction (two patches per linker), such a constraint could be practically implemented in a variety of ways for linker systems. Linkers can be ions or molecules, which possess intrinsic limitations on bond formation, or size asymmetry between linker and primary particle may naturally enforce such a constraint. For the primary particles, an interesting future direction would be to explore the cases where $n_{\mathcal{P}}>6$ in order to further reduce the presence of microscopic valence restriction in the model. (Though a 6-patch model has been shown to be experimentally realizable,~\cite{exp_patchy_coll} such architectures remain challenging to fabricate.) As $n_{\mathcal{P}}$ increases, the results derived in the Appendix for arbitrary patch number indicate that $\Gamma_{\text{L}}$ would be slightly reduced, tending towards $\Gamma_{\text{L}}=0.5$ as $n_{\mathcal{P}} \rightarrow \infty$. By contrast, on the high-linker side, $\Gamma_{\text{U}}$ will continue to grow with $n_{\mathcal{P}}$ as indicated by Eqn.~\ref{eqn:upper_bonding}. The manner in which such predictions are modified by real-world effects, such as looping and/or steric repulsion is an interesting question for future investigations. Finally, a promising avenue for future work is to explore how introducing correlations to the bonding, for instance via distance-dependent repulsions between like particles, might be able to further enhance equilibrium gel formation by possibly reducing loop formation or by more uniformly distributing the linker among the primary particles. 

\section{Acknowledgments}
This work was partially supported by the National Science Foundation (1247945) and the Welch Foundation (F-1696 and F-1848). We acknowledge the Texas Advanced Computing Center (TACC) at The University of Texas at Austin for providing HPC resources.

\setcounter{figure}{0}
\setcounter{equation}{0}
\renewcommand\thefigure{A\arabic{figure}}
\renewcommand{\thesection}{\thepart .\arabic{section}}
\renewcommand\theequation{A\arabic{equation}}

\renewcommand{\thesubsection}{\arabic{subsection}}

\section*{Appendix: Phase coexistence and percolation}
\label{sec:appendix}

\subsection{Computing Phase Coexistence}

Here we describe out strategy for the explicit phase coexistence calculations. One possibility is to solve the non-linear equations equating the chemical potential of each species in the two phases as well as two phases pressures. However, a more attractive alternative is to follow the standard textbook thermodynamic minimization of the Helmholtz free energy of two arbitrary phases \(k={1,2}\) 
\begin{equation} \label{eqn:extensive_two_phase_free_energy}
A_{1,2}\equiv A_{1}+A_{2}
\end{equation}
where \(A_{k}=A(N_{\mathcal{P},k},N_{\mathcal{L},k},V_{k},T)\) is the Helmholtz free energy of phase \(k\) constrained to have \(N_{\mathcal{P},k}\) and \(N_{\mathcal{L},k}\) particles of species \(\mathcal{P}\) and \(\mathcal{L}\) respectively in volume \(V_{k}\).~\cite{thermo_stat_mech} Finding coexisting phases amounts to relaxing the constraints and minimizing the \(A_{1,2}\) subject to particle number [\(N_{\mathcal{P}},N_{\mathcal{L}}\)] and volume preservation [\(V\)] via
\begin{equation} \label{eqn:extensive_free_energy_minimization}
\begin{split}
& A_{1,2}^{(c)} \equiv \underset{{\{N_{k},V_{k}\}}}{\text{min}} A_{1,2} \\
& V_{1}+V_{2}=V \\
& N_{\mathcal{P},1}+N_{\mathcal{P},2}=N_{\mathcal{P}} \\
& N_{\mathcal{L},1}+N_{\mathcal{L},2}=N_{\mathcal{L}}
\end{split}
\end{equation}
where \(A_{1,2}^{(c)}\equiv A_{1,2}^{(c)}(N_{\mathcal{P}},N_{\mathcal{L}},V)\) is Helmholtz free energy of the coexisting (c) phase system, each phase of which is described by the optimized variables [\(N_{\mathcal{P},k}^{(c)},N_{\mathcal{L},k}^{(c)},V_{k}^{(c)}\)] which depend on the state point control variables [\(N_{\mathcal{P}},N_{\mathcal{L}},V,T\)]. This approach is superior as the free energy landscape generally possesses at most a few minima (i.e., when the homogeneous fluid is metastable with respect to phase coexistence) and thus, in principle, downhill minimizers can easily be employed without strong dependence on the starting guess. 

In practice though, the problem must be recast in terms of intensive variables. In this scenario, one instead prescribes (\(\rho,x_{\mathcal{P}},T\)), where these variables now should be regarded as simply setting how many particles of each species are contained in a volume through intensive variable alternatives, regardless of whether the system is deemed single- or two-phase after minimization. Using the definitions: \(\rho_{k}\equiv (N_{\mathcal{P},k}+N_{\mathcal{L},k})/V_{k}\), \(\theta_{k}\equiv V_{k}/V\), \(x_{\mathcal{P},k}\equiv N_{\mathcal{P},k}/N_{k}\) and \(a_{k}\equiv A_{k}/N_{k}= a(\rho_{k},x_{\mathcal{P},k},T)\) we arrive at the following intensive formulation of Eqn. \ref{eqn:extensive_two_phase_free_energy} 
\begin{equation} \label{eqn:intensive_two_phase_free_energy}
a_{1,2}\equiv \theta_{1}\bigg(\dfrac{\rho_{1}}{\rho}\bigg)a_{1}+\theta_{2}\bigg(\dfrac{\rho_{2}}{\rho}\bigg)a_{2}
\end{equation}
The constrained, intensive two phase Helmholtz free energy minimization--equivalent to Eqn. \ref{eqn:extensive_free_energy_minimization}--then follows as
\begin{equation} \label{eqn:intensive_free_energy_minimization}
\begin{split}
& a_{1,2}^{(c)} \equiv \underset{{\{\rho_{k},x_{a,k},\theta_{k}\}}}{\text{min}} a_{1,2} \\
&0\leq x_{\mathcal{P},1},x_{\mathcal{P},2}\leq1 \\
&0\leq \theta_{1},\theta_{2}\leq1 \\
&\theta_{1}+\theta_{2}=1 \\
& x_{\mathcal{P},1}\rho_{1}\theta_{1}+x_{\mathcal{P},2}\rho_{2}\theta_{2}=x_{\mathcal{P}}\rho \\
& (1-x_{\mathcal{P},1})\rho_{1}\theta_{1}+(1-x_{\mathcal{P},2})\rho_{2}\theta_{2}=(1-x_{\mathcal{P}})\rho
\end{split}
\end{equation}
where \(a_{1,2}^{(c)}\equiv a_{1,2}^{(c)}(\rho,x_{\mathcal{P}},T)\) is the Helmholtz free energy per particle of the coexisting phase system, each phase of which is described by the optimized variables [\(\rho_{k}^{(c)},x_{\mathcal{P},k}^{(c)},\theta_{k}^{(c)}\)] which depend on the state point control variables [\(\rho,x_{\mathcal{P}},T\)].

As a first simplification, three variables were eliminated algebraically via the last three linear constraints in Eqn.~\ref{eqn:intensive_free_energy_minimization} leaving only [\(\rho_{1},x_{\mathcal{P},1},\rho_{2}\)] for minimization. Minimization was then accomplished via a Simplex scheme which very rapidly finds either the homogeneous phase or a phase-separated state.

\subsection{Derivation of Analytical Percolation Results}

From the percolation theory established in Section~\ref{subsec:methods_percolation}, \emph{general} analytical predictions can be made for a binary system (\(q={\mathcal{P},\mathcal{L}}\)) of \(N_{q}\) particles of species \(q\) with \(n_{q}\) sites that can only bind unlike particles sites. We assume that \(n_{\mathcal{P}}>n_{\mathcal{L}}\) and thus regard the \(\mathcal{L}\) species as the linker and the \(\mathcal{P}\) species the primary particle. The maximal number of bonds can form when \(N_{\mathcal{P}}n_{\mathcal{P}}=N_{\mathcal{L}}n_{\mathcal{L}}\) which after adopting the definition of the linker to primary ratio, \(\Gamma\equiv N_{\mathcal{L}}/N_{\mathcal{P}}\), yields the zero temperature~\footnote{The criterion used only finds the conditions that would \emph{permit} the maximal number of bonds and says nothing about whether this would be realized at finite T. Only at zero T is this statement exact; however, for the low temperatures relevant to our system deviations should be unimportant.} condition of maximal bonding
\begin{equation} \label{eqn:max_bonding}
\Gamma_{\text{M}}\equiv n_{\mathcal{P}}/n_{\mathcal{L}}
\end{equation}
For \(\Gamma<\Gamma_{\text{M}}\) the linker is the limiting ``reagent'' while for \(\Gamma>\Gamma_{\text{M}}\) it is the primary. 

At \(T=0\) this maximally bonded mixture trivially satisfies the percolation threshold criterion (Eqn.~\ref{eqn:percolation_criterion}) since \(p_{\mathcal{P} \rightarrow \mathcal{L}}=1\) and \(p_{\mathcal{L} \rightarrow \mathcal{P}}=1\); however, there exists a lower \(\Gamma_{\text{L}}<\) and upper \(\Gamma_{\text{U}}\) satisfying \(\Gamma_{\text{L}}<\Gamma_{\text{M}}<\Gamma_{\text{U}}\) such that below the former and above the latter percolation at \(T=0\) (and thus all finite temperature) is no longer possible. For \(\Gamma<\Gamma_{\text{M}}\), all linker sites are bonded, \(p_{\mathcal{L} \rightarrow \mathcal{P}}=1\), and the best \(p_{\mathcal{P} \rightarrow \mathcal{L}}\) (achieved at \(T=0\)) is the total number of \(\mathcal{L}\) sites per the total number of \(\mathcal{P}\) sites, \(p_{\mathcal{P} \rightarrow \mathcal{L}}=n_{\mathcal{L}}N_{\mathcal{L}}/(n_{\mathcal{P}}N_{\mathcal{P}})=n_{\mathcal{L}}\Gamma/n_{\mathcal{P}}\). Using these relations in Equation~\ref{eqn:percolation_criterion} yields
\begin{equation} \label{eqn:lower_bonding}
\Gamma_{\text{L}}\equiv \dfrac{n_{\mathcal{P}}}{n_{\mathcal{L}}(n_{\mathcal{P}}-1)(n_{\mathcal{L}}-1)}
\end{equation}
Similar arguments allow for an analytical calculation of \(\Gamma_{\text{U}}\). For \(\Gamma>\Gamma_{\text{M}}\), all primary sites are bonded, \(p_{\mathcal{P} \rightarrow \mathcal{L}}=1\), and the best \(p_{\mathcal{L} \rightarrow \mathcal{P}}\) is the total number of \(\mathcal{P}\) sites per the total number of \(\mathcal{L}\) sites, \(p_{\mathcal{L} \rightarrow \mathcal{P}}=n_{\mathcal{P}}N_{\mathcal{P}}/(n_{\mathcal{L}}N_{\mathcal{L}})=n_{\mathcal{P}}/(n_{\mathcal{L}}\Gamma)\). Using these relations in Equation~\ref{eqn:percolation_criterion} yields
\begin{equation} \label{eqn:upper_bonding}
\Gamma_{\text{U}}\equiv \dfrac{n_{\mathcal{P}}(n_{\mathcal{P}}-1)(n_{\mathcal{L}}-1)}{n_{\mathcal{L}}}
\end{equation}

Finally, also of interest are \(\Gamma\) values below and above \(\Gamma_{\text{M}}\) that have a matched ``propensity'' for percolation as quantified by \((n_{\mathcal{P}}-1)p_{\mathcal{P} \rightarrow \mathcal{L}}(n_{\mathcal{L}}-1)p_{\mathcal{L} \rightarrow \mathcal{P}}\). Using the same rules for the \(T=0\) probabilities, \(p_{\mathcal{L} \rightarrow \mathcal{P}}\) and \(p_{\mathcal{P} \rightarrow \mathcal{L}}\), discussed above and defining \(\Gamma_{+}\) and \(\Gamma_{-}\) as linker to primary ratios below (primary rich) and above (linker rich) \(\Gamma_{\text{M}}\) respectively we find
\begin{equation} \label{eqn:equi_percolation}
\Gamma_{+}=\Gamma_{-}^{-1} (\dfrac{n_{\mathcal{P}}}{n_{\mathcal{L}}})^{2}
\end{equation}
Thus, given a linker poor mixture we can predict a linker rich analog (or vice-versa) with a similar percolation ``propensity''.



%


\end{document}